\documentclass[10pt,letterpaper,twocolumn,conference]{IEEEtran}
\usepackage[latin9]{inputenc}
\usepackage{amsmath}
\usepackage{amssymb}
\usepackage{stackrel}
\usepackage{graphicx}

\makeatletter

\pdfpageheight\paperheight
\pdfpagewidth\paperwidth

\usepackage[nospace]{cite}
\usepackage[format=hang,justification=justified,singlelinecheck=false,font=small]{caption}
\usepackage[table]{xcolor}
\definecolor{GrayTable}{rgb}{.9,.9,.9}

\makeatother

\begin{document}

\title{On the Transport Capability of LAN Cables\linebreak{}
in All-Analog MIMO-RoC Fronthaul}

\author{\IEEEauthorblockN{Syed Hassan Raza Naqvi, Andrea Matera, Lorenzo Combi and Umberto Spagnolini} 

\IEEEauthorblockA{Dipartimento di Elettronica, Informazione e Bioingegneria (DEIB), Politecnico di Milano, Italy}

E-mails: \{syedhassan.naqvi, andrea.matera, lorenzo.combi, umberto.spagnolini\}@polimi.it 
}
\maketitle
\begin{abstract}
Centralized Radio Access Network (C-RAN) architecture is the only
viable solution to handle the complex interference scenario generated
by massive antennas and small cells deployment as required by next
generation (5G) mobile networks. In conventional C-RAN, the fronthaul
links used to exchange the signal between Base Band Units (BBUs) and
Remote Antenna Units (RAUs) are based on digital baseband (BB) signals
over optical fibers due to the huge bandwidth required. In this paper
we evaluate the transport capability of copper-based all-analog fronthaul
architecture called Radio over Copper (RoC) that leverages on the
pre-existing LAN cables that are already deployed in buildings and
enterprises. In particular, the main contribution of the paper is
to evaluate the number of independent BB signals for multiple antennas
system that can be transported over multi-pair Cat-5/6/7 cables under
a predefined fronthauling transparency condition in terms of maximum
BB signal degradation. The MIMO-RoC proves to be a complementary solution
to optical fiber for the last 200m toward the RAUs, mostly to reuse
the existing LAN cables and to power-supply the RAUs over the same
cable.
\end{abstract}


\section{Introduction\label{sec:Introduction}}

The radio access network (RAN) paradigm is completely changing in
next generation (5G and beyond) mobile systems due to the ever-growing
traffic demand, calling for the fulfillment of strict requirements
in terms of throughput, mobility and latency. Pervasive deployment
of a large number of antennas appears to be the only viable solution
to meet such requirements \cite{Boccardi-Heath-etal_2014}, though
introducing complicated interference scenarios. In this context, Centralized
RAN (C-RAN) \cite{Checko-Christiansen-etal_2015} addresses the challenging
interference management task through centralized processing to handle
a massive number of antennas/cells, taking benefits from their mutual
cooperation for interference mitigation. Key enabler for C-RAN is
the co-location of baseband units (BBUs) in so-called BBU pools, that
allows centralized processing providing remarkable benefits in terms
of programmability, scalability and cost reduction. The antennas,
with all the radio-frequency (RF) functionalities, are hosted at the
remote antenna units (RAUs) while modulation/demodulation and precoding
are performed at the BBUs. The fronthaul link between RAUs and BBUs
is conventionally designed to exchange digital in-phase and quadrature
(I/Q) signals streaming according to the CPRI protocol \cite{CPRI:specs},
demanding analog/digital conversion at the RAU. 

\begin{figure}[t]
\noindent \begin{centering}
\includegraphics[width=1\columnwidth]{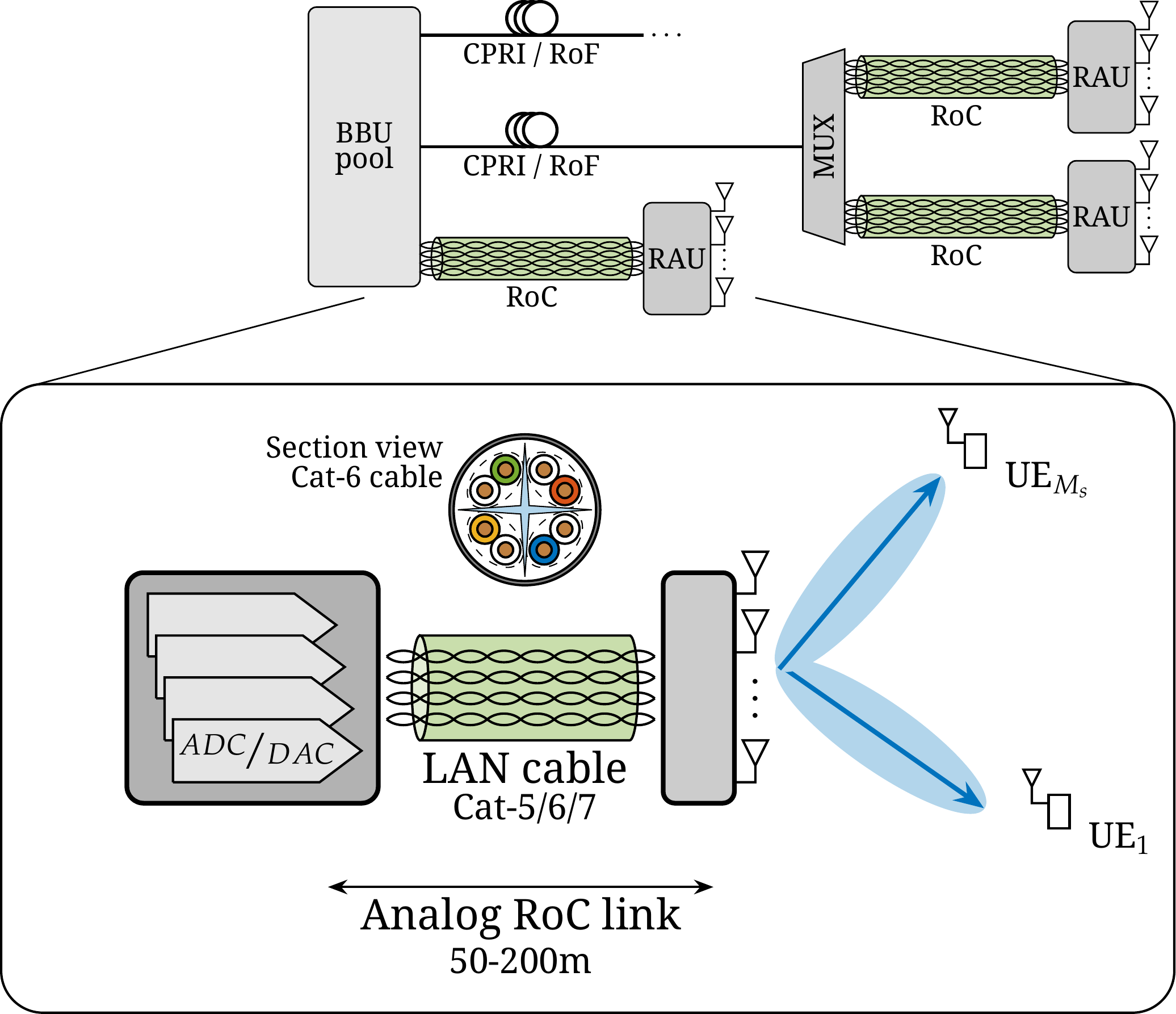}
\par\end{centering}
\noindent \begin{centering}
\caption{C-RAN architecture with joint deployment of fiber and copper \label{fig:CRAN_vision}}
\par\end{centering}
\vspace{-20pt}
\end{figure}

The expected increase in RF signal bandwidth and the massive number
of antennas call into question the effectiveness of this digital I/Q
streaming, that would introduce a bandwidth expansion of RF signals
over fronthauling that can be as severe as 30x. Compression of digital
I/Q signals \cite{Compression:Nanba} or RAN functional splits \cite{Bartelt:FunctSplit}
are not enough for bandwidth reduction. In all-analog fronthauling
the RAUs directly relay the analog BB signals (possibly after frequency
translation) to/from the BBUs (or any midway equipment) thus avoiding
any bandwidth expansion, minimizing the latency (that is just the
propagation) and hardware cost while improving energy efficiency \cite{Gambini-Spagnolini_2013,wake:RoFdesign,chung:rofFronthaul}.
Analog radio over fiber (RoF) provides an effective example of analog
fronthauling, due to its capability to carry several Gbit/s in terms
of equivalent data-rate \cite{chung:rofFronthaul,wake:RoFdesign}.
However, RoF infrastructure would require the deployment of a large-scale
and pervasive fiber optic infrastructure whose cost can be excessive.
Even if RoF could be based on passive optical networks (PON), the
RAUs still need the power supply for optical sources and electronics,
and this could make the economical benefits of using PON questionable,
mostly for the last 100-200m of fronthauling. These issues can be
dealt with by a hybrid fiber-cable architecture (Fig. \ref{fig:CRAN_vision}),
as foreseen in \cite{effenberger2016future}.

Radio over Copper (RoC) \cite{Gambini-Spagnolini_2013,RadioDots2014}
is a complementary/alternative technology for fronthauling that makes
use of LAN cables which are nowadays already largely deployed in buildings
and enterprises, thus cutting the costs of the deployment of a brand-new
network infrastructure, mostly if considering indoor coverage. Here
we analyze the potential of RoC provided by multi-pair LAN cables
containing 4 twisted pairs bounded together to provide at least 4
separated cable-pairs (or space) channels with a bandwidth of up to
1GHz/pair (depending on the cable type). Cat-5/6/7 cables enable the
design of all-analog and low-cost fronthauling (at least for the last
100-200m) capable of supporting a large number of independent radio
access channels (possibly corresponding to different radio access
technologies, RATs) or a massive number of antennas at each RAU. The
equivalent data-rate over copper-cable systems is large enough to
make RoC extend fiber optics coverage.

Differently from \cite{Gambini-Spagnolini_2013,Medeiros-Huang-etal_2016},
where each BBU-RAU connection is provided by a single twisted pair,
the use of multi-pair LAN cables introduces an additional multiplexing
dimension for the wired fronthaul channel as at least 4 mutually interfering
channels are available for each frequency band (more channels if considering
the phantom-modes of cables). The overall channel between the BBU
and the end-user is therefore modeled as the cascade of a MIMO radio
channel and a MIMO cable channel, where the signal to/from each antenna
needs to be properly mapped onto the available cable-frequency resources
within the LAN cable, defined here as space-frequency domain of the
cable. In this paper the Cat-5/6/7 cables characteristics are revised
to delineate the link budget for the cascade of air and copper links.
The link-budget helps to define the maximum number of independent
RAT channels (or independent antennas in every RAU) that can be allocated
on different cables/frequency bands taking into account the maximum
power spectral density over cables to avoid extra-cable interference. 

\textit{Contributions: }The contributions of this paper are three-fold:
\textit{i)} we propose an all-analog MIMO-RoC fronthauling based on
LAN cables supporting massive numbers of antennas over independent
copper-links, \textit{ii) }we show the potential of Cat-5/6/7 cables
by evaluating the number of independent RAT channels that different
cable types and lengths can accommodate without any air-link degradation,
\textsl{iii)} we propose a dynamic power allocation for MIMO-RoC that
enables to minimize any cable cross-talk.

\emph{Notation: }Bold upper- and lower-case letters describe matrices
and column vectors. $\left[\mathbf{A}\right]_{ij}=a_{ij}$ denotes
the \emph{ij}-th element of matrix $\mathbf{A}.$ Letters $\mathbb{R},\mathbb{C}$
refer to real and complex numbers, respectively. We denote matrix
inversion, transposition and conjugate transposition as $\left(\cdot\right)^{-1},\left(\cdot\right)^{T},\left(\cdot\right)^{H}$
and matrix $\mathbf{I}$ is an identity matrix of appropriate size. 

\textit{Organization}: Section \ref{sec:System-Model-2} introduces
the system model for the radio scenario and the MIMO-RoC fronthauling,
considering the combined effects of air and cable interference. LAN
cable characteristics are summarized in Section \ref{sec:LAN-Cables-for},
while in Section \ref{sec:Spectrum-Balancing-in} power resource allocation
is presented. Numerical results are in Section \ref{sec:Numerical Results}
and concluding remarks and future works are highlighted in Section
\ref{sec:Conclusion}. 

\section{System Model and Parameters \label{sec:System-Model-2}}

\begin{figure}
\vspace{3pt}

\noindent \begin{centering}
\includegraphics[width=1\columnwidth]{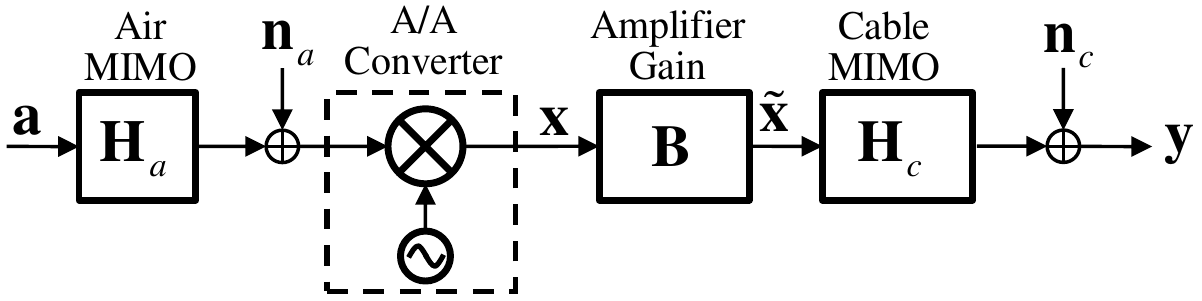}
\par\end{centering}
\noindent \begin{centering}
\caption{Block diagram of MIMO-RoC.\label{fig:Block-diagram-of}}
\par\end{centering}
\vspace{-10pt}
\end{figure}

The system model for the proposed MIMO-RoC fronthauling is shown in
Fig. \ref{fig:Block-diagram-of} for uplink transmission (downlink
would be similar, not covered here), where $N_{a}$ antennas at the
RAU are connected to the BBU by a LAN cable with $N_{c}=4$ twisted
pairs to serve $N_{u}$ users. To simplify the reasoning here, users
are equipped with one antenna each, but any generalization is straightforward.
Each RAU relays towards the BBU (or any other mid-way equipment) the
analog signals received from the users after a proper analog-to-analog
(A/A) conversion consisting in frequency downconversion of RF signal
to match the cable frequency band and an amplifier to be detailed
later. 

The signal $\mathbf{x}\in\mathbb{C}^{N_{a}}$ received at the antenna
array (after downconversion) is modeled as flat-fading (e.g., it corresponds
to one subcarrier of OFDM/OFDMA modulation)
\begin{equation}
\mathbf{x}=\mathbf{H}_{a}\mathbf{a}+\mathbf{n}_{a},\label{eq:Input_at_RRU}
\end{equation}
where $\mathbf{H}_{a}\in\mathbb{C}^{N_{a}\times N_{u}}$ is the air-link
channel matrix from the $N_{u}$ users to the $N_{a}$ antennas, $\mathbf{a}\in\mathbb{C}^{N_{u}}$
is the users signal, and $\mathbf{n}_{a}\sim\mathcal{CN}(\mathbf{0},\sigma_{a}^{2}\mathbf{I})$
is the air-link noise at the antenna array with power $\sigma_{a}^{2}$.
The signal $\mathbf{y}\in\mathbb{C}^{N_{a}}$ received at the BBU
after A/A is

\begin{align}
\mathbf{y} & =\mathbf{H}_{c}\mathbf{B}\mathbf{x}+\mathbf{n}_{c},\label{eq:BBU signal}
\end{align}
where $\mathbf{H}_{c}\in\mathbb{C}^{N_{a}\times N_{a}}$ is the space-frequency
MIMO cable channel accounting for the frequency-dependent insertion
loss (IL, on the main diagonal elements $[\mathbf{H}_{c}]_{i,i}$)
and for the frequency-dependent far-end crosstalk (FEXT, on the off-diagonal
terms $[\mathbf{H}_{c}]_{i,j}$). The $N_{a}$ signals are channelized
in frequency division multiplexing (FDM) over $N_{f}$ cable bands
with $N_{c}$ pairs/ea., the cable channel is block-diagonal to guarantee
the orthogonality over the cable-bands:

\begin{equation}
\mathbf{H}_{c}=\text{diag}\big[\mathbf{H}_{c,1},\mathbf{H}_{c,2},\dots,\mathbf{H}_{c,N_{f}}\big].\label{eq:ref_channel_diag}
\end{equation}
Cable channel at the $k$-th frequency band $\mathbf{H}_{c,k}\in\mathbb{C}^{N_{c}\times N_{c}}$
contains the insertion loss and crosstalk elements for the $N_{c}$
spatial cable channels. The total bandwidth available over the $N_{f}\cdot N_{c}$
space-frequency cable channels should be $N_{f}N_{c}\geq N_{a}$ for
a given signal to interference plus noise ratio (SINR) degradation
of the cable (Sect. \ref{subsec:Cable-Resources-Assignment}).

In order to cope with cable crosstalk without any digital signal processing,
the only degrees of freedom are the antenna mapping onto cable and
the design of the amplification for each copper/frequency-link. $\mathbf{B}$
is the amplifier gain matrix that is designed to minimize the cable
crosstalk effect by adjusting the transmitted power over each space-frequency
cable channel as detailed in Sect.\ref{sec:Spectrum-Balancing-in}.
The additive noise introduced by the cable is $\mathbf{n}_{c}\sim\mathcal{CN}(\mathbf{0},\sigma_{c}^{2}\mathbf{I})$
and (\ref{eq:BBU signal}) becomes
\begin{equation}
\mathbf{y}=\mathbf{H}_{c}\tilde{\mathbf{x}}+\mathbf{n}_{c},\label{eq:BBU signal v2}
\end{equation}
where $\tilde{\mathbf{x}}$ contains the signals transmitted from
RAU to BBU over the $N_{c}$ twisted pairs and $N_{f}$ cable frequency
bands. 

For LAN cables $N_{c}=4$, and the total bandwidth depends on cable
length and type. Even if their bandwidth can be as high as approx.
1 GHz, here the analysis is only for the first 500 MHz for the availability
of experimental measurements and models. Moreover, the analysis is
for multiple antennas LTE system with 20-MHz bandwidth that uses a
global bandwidth of 22 MHz comprehensive of 10\% of guard band overhead
for FDM over cables. Even if the cable channels are frequency-dependent,
for the sake of simplicity in numerical analysis, the cable channel
matrix $\mathbf{H}_{c}$ is approximated as constant within each 22-MHz
band.

\section{LAN Cables for MIMO-RoC \label{sec:LAN-Cables-for}}

The fronthaul capacity of the RoC architecture in \cite{Gambini-Spagnolini_2013}
can be greatly enhanced through MIMO-RoC, that efficiently exploits
cables with multiple twisted pairs as the 4-pairs of LAN cables detailed
below. 

\subsection{Cables Characteristic}

Standard LAN cables (Cat-5/6/7) are considered here for MIMO-RoC.
Cat-5 cables are unshielded twisted pair (UTP) cables, commonly deployed
for computer networks (e.g., Gigabit Ethernet) and their performances
are mostly affected by the system noise. Cat-6 cables are high-grade
UTP cables with additional foil underneath the cable jacket, in which
better noise and interference immunity is achieved by increasing the
twists density ($>2$ twists/cm for Cat-6, compared to 1.5 twists/cm
for Cat-5). Cat-7 cables offer lower signal attenuation and reduced
intra-cable interference through extensive shielding and foiling over
each individual twisted pair, and therefore can be used in a noisy
environment, or to remarkably increase the transport capability with
respect to Cat-5 cables.

\begin{figure}[th]
\vspace{2pt}

\noindent \begin{centering}
\includegraphics[width=1\columnwidth]{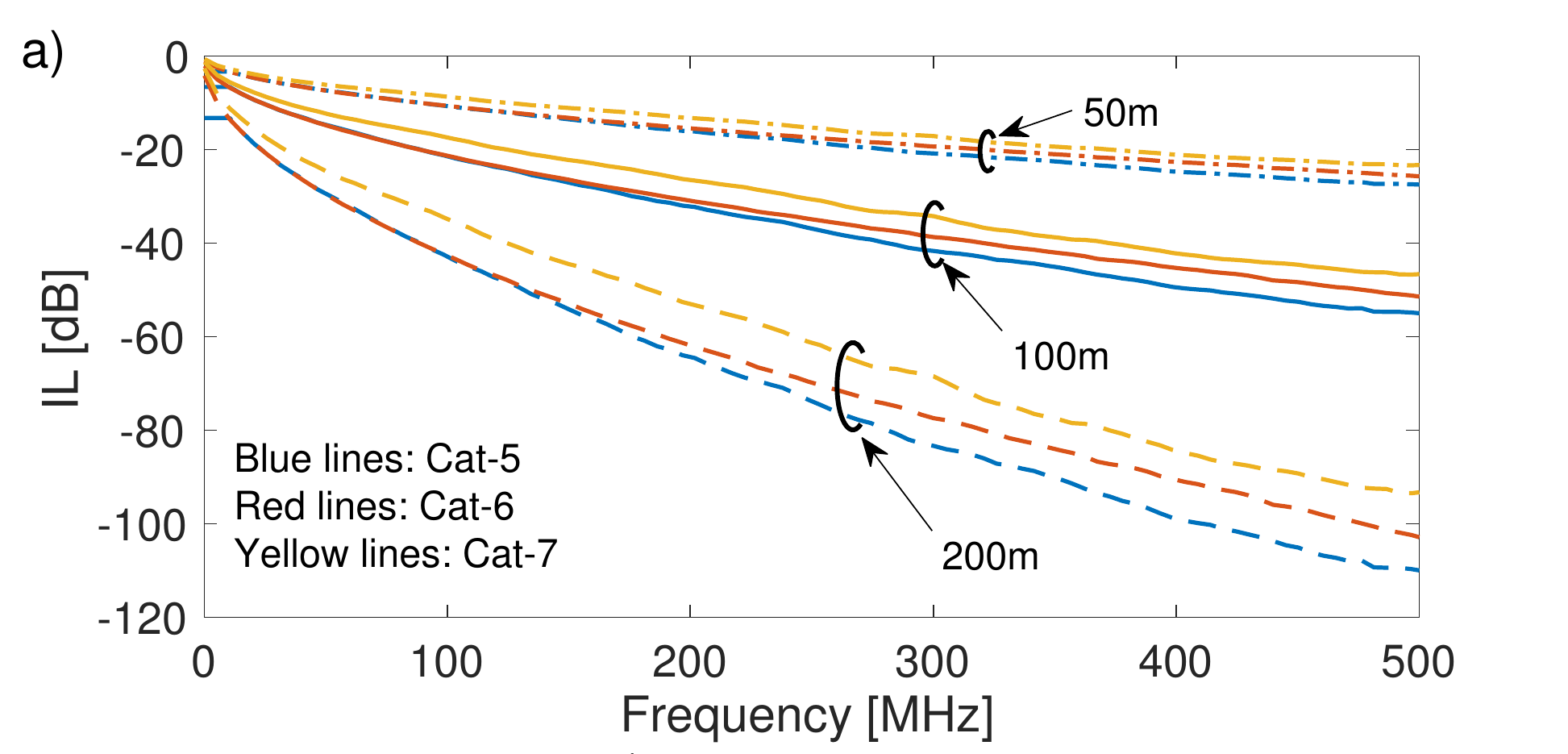}
\par\end{centering}
\includegraphics[width=1\columnwidth]{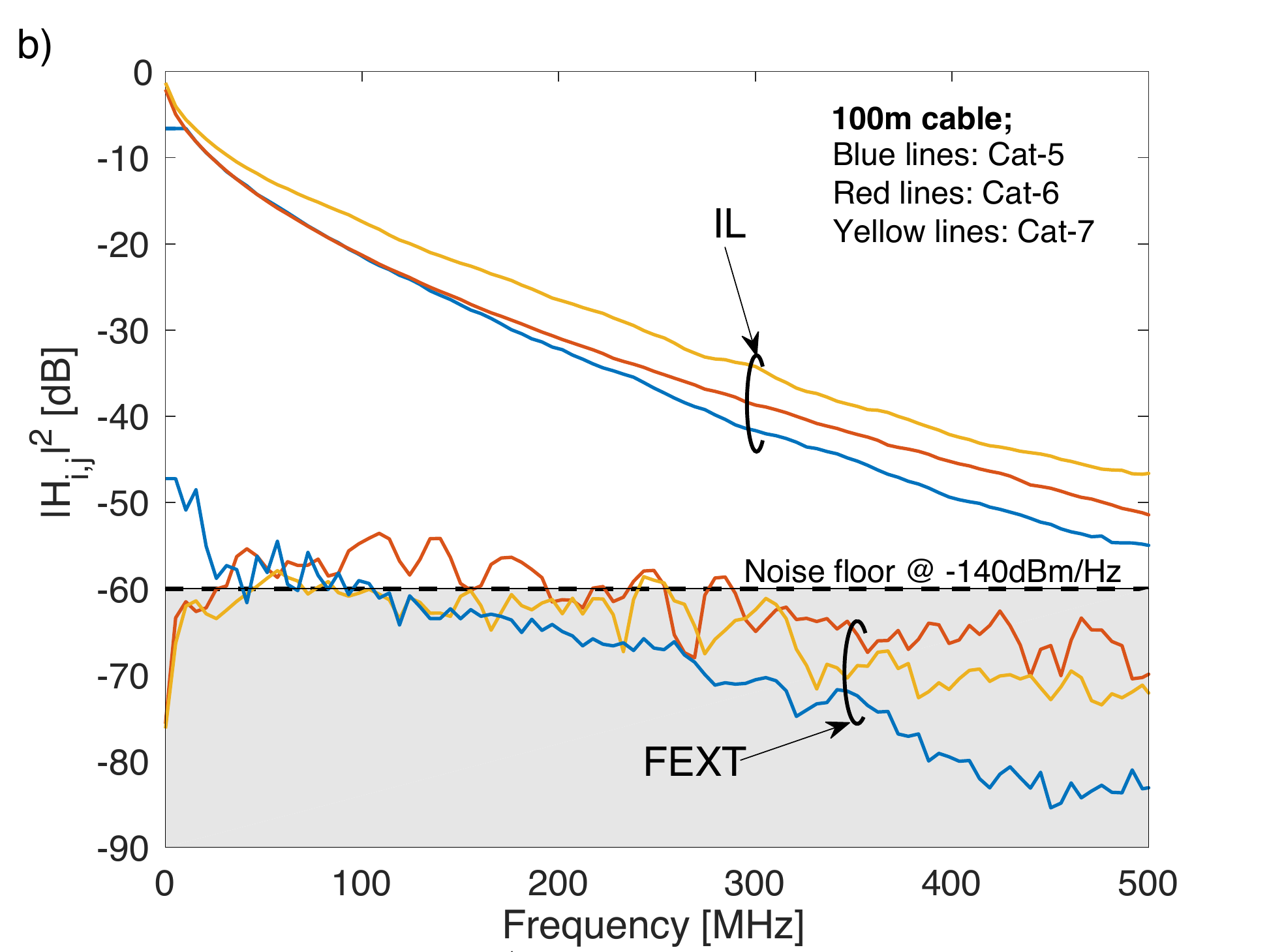}

\vspace{-5pt}

\noindent \begin{centering}
\caption{Cable characteristics: a) insertion loss vs frequency for 50m, 100m
and 200m and Cat-5/6/7 cables; b) cable gain (IL) and crosstalk (FEXT)
vs frequency for 100-m cables (noise-floor is scaled from standard
cable-design parameters) \label{fig:LAN-Cable-Characteristics}}
\par\end{centering}
\vspace{-5pt}
\end{figure}

The transmission bandwidth over each copper link is typically considered
up to 212 MHz for 100-m cables (e.g., G.fast as next generation DSL),
but it can reach as much as 1 GHz for LAN cables, especially for shorter
distances ($\leq50$ m). Cable lengths can vary in a range between
few meters up to several hundreds of meters, but higher insertion
losses are associated with longer cables, thus reducing the practically
available transmission bandwidth as shown in Fig. \ref{fig:LAN-Cable-Characteristics}.
The analysis is limited here to 50-m, 100-m and 200-m LAN cables over
maximum bandwidth of 500 MHz (limit due to the reliability of available
measurements). Notice that, even thought the total available bandwidth
of each pair is $B_{c,max}=500\text{ MHz}$, the useful bandwidth
for signal transmission $B_{c}$ is reduced for the joint effect of
cable attenuation and crosstalk and $B_{c}\leq B_{c,max}$.

Fig.\ref{fig:LAN-Cable-Characteristics} shows the characteristics
of the 3 categories of cables in terms of average (over pairs) insertion
loss versus frequency for varying cable length (Fig.\ref{fig:LAN-Cable-Characteristics}a),
and in terms of average IL and FEXT for 100-m length for varying Cat-type
(Fig.\ref{fig:LAN-Cable-Characteristics}b). In Fig.\ref{fig:LAN-Cable-Characteristics}b
the noise-floor is highlighted assuming that its power spectral density
(typ. is $-140$ dBm/Hz) is normalized to the maximum spectral mask
of signal ($-80$ dBm/Hz). Cables have a low-pass characteristic that
increases with cable length as shown in Fig.\ref{fig:LAN-Cable-Characteristics}a.
For the scope of having a transparent RoC, it is meaningful to derive
from Fig.\ref{fig:LAN-Cable-Characteristics} the SINR versus frequency
that is illustrated in Fig.\ref{fig:Cat6_characteristics} for Cat-6
cable and varying cable lengths (50, 100 and 200 m). It can be noticed
from Fig.\ref{fig:Cat6_characteristics} that in case of longer cables
(i.e., 200 m) the noise dominates over FEXT limiting its usage up
to 200 MHz. 

\subsection{Cable Resources Analysis\label{subsec:Cable-Resources-Assignment}}

\begin{figure}[t]
\includegraphics[width=1\columnwidth]{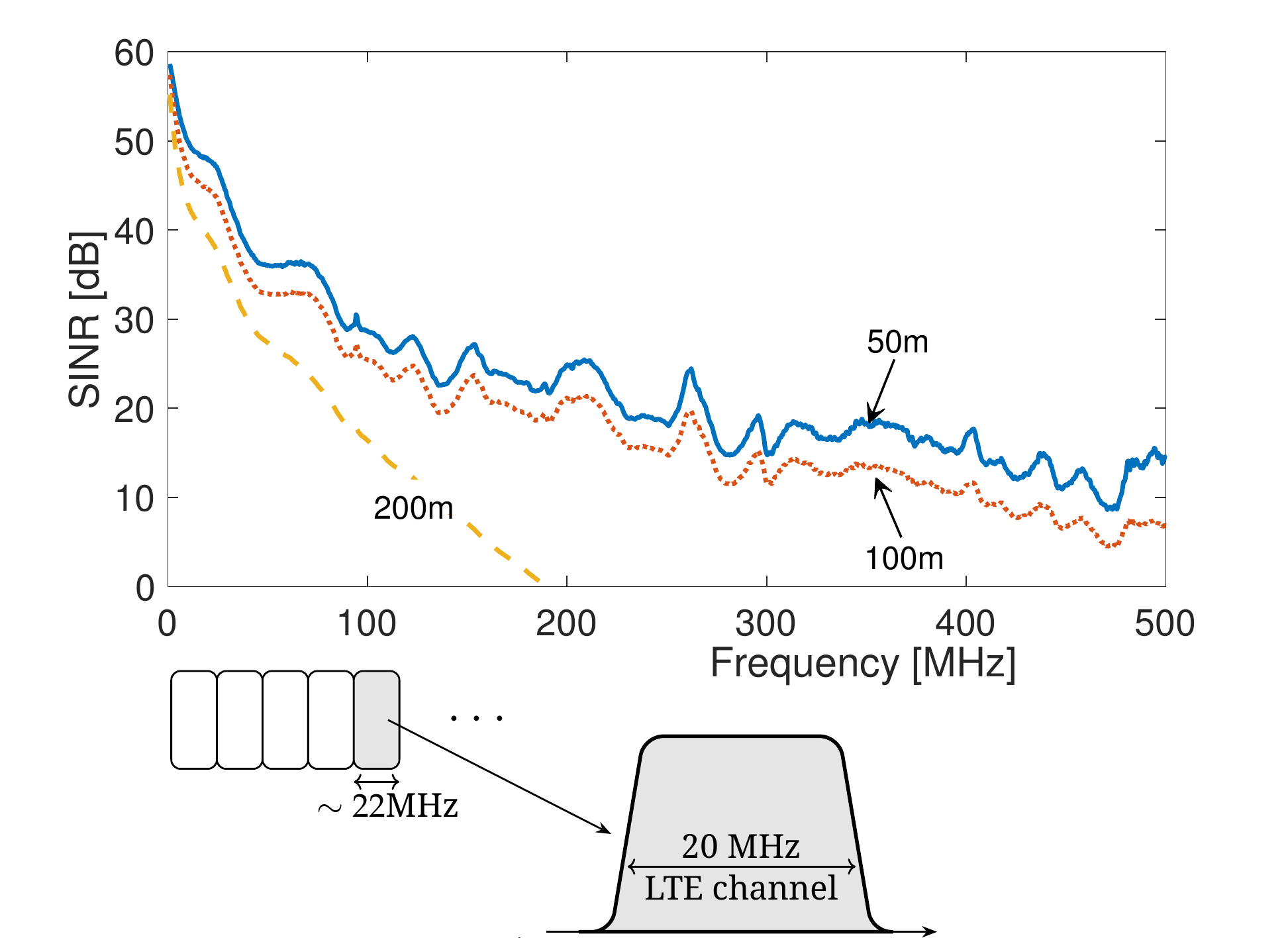}

\caption{Bandwidth of LTE signals and SINR vs frequency for Cat-6 cable, 100m
length and its extrapolation to 50m, 200m \label{fig:Cat6_characteristics}}

\vspace{-12pt}
\end{figure}

In the proposed MIMO-RoC fronthauling architecture, the signal from
each antenna is mapped over one of the different 22-MHz bands of LAN
cable, as shown in Fig.\ref{fig:Cat6_characteristics}. Given the
bandwidth of each air-link $B_{a}$ (that in the numerical analysis
is $B_{a}=22$ MHz), the bandwidth to transport over cable the overall
$N_{a}$ antennas is $N_{a}B_{a}$, hence the cable transport capability
must be 
\begin{equation}
\sum_{\ell=1}^{N_{c}}B_{c,\ell}\geq N_{a}B_{a},\label{eq:Cable_Capacity}
\end{equation}
where $B_{c,\ell}$ is the useful transmission bandwidth on the $\ell$-th
twisted pair of the $N_{c}$ in a cable subject to a minimal SINR
level induced by the cable from Fig.\ref{fig:Cat6_characteristics}.
From (\ref{eq:Cable_Capacity}), $N_{a}\leq N_{a,max}$ and thus 
\begin{equation}
N_{a,max}=\sum_{\ell=1}^{N_{c}}\left\lfloor \frac{B_{c,\ell}}{B_{a}}\right\rfloor \label{eq:Cable_Capacity-1}
\end{equation}
is the maximum number of independent RAT channels (antennas) that
can be allocated onto a LAN cable. 

\section{Power Allocation in MIMO-RoC \label{sec:Spectrum-Balancing-in}}

The main limitation in LAN cables is the crosstalk among the twisted
pairs. Crosstalk cancellation is the obvious solution to maximize
the throughput over the copper, but it requires the digitalization
and the corresponding digital signal processing at the A/A converter,
that is too energy and latency consuming compared to the all-analog
relaying (not to mention the cost). The aim of the power allocation
in MIMO-RoC is to optimize the power of the transmit signal $\tilde{\mathbf{x}}$
over the space-frequency copper channels such that the crosstalk among
the cable pairs is minimized. Since copper channel gains (IL and FEXT)
are time-invariant compared to the air-links, the power allosubcation
problem consists in optimizing the gains of the diagonal matrix $\mathbf{B}$
that scales the power of the input signal $\mathbf{x}$ prior to transmission
over the cable. This can be solved by adapting the optimum spectrum
balancing (OSB) algorithm \cite{8,9} to RoC.

The SINR at BBU follows from (\ref{eq:BBU signal v2}) and it can
be represented for line $n$ and $k$-th subcarrier as
\begin{equation}
\text{SINR}_{k}^{n}=\frac{\left|h_{k}^{n,n}\right|^{2}p_{k}^{n}}{\underset{m\neq n}{\sum}\left|h_{k}^{n,m}\right|^{2}p_{k}^{m}+\sigma_{k}^{n}},
\end{equation}
where $h_{k}^{n,m}=[\mathbf{H}_{c,k}]_{n,m}$ is the channel gain
from cable pair $m$ towards $n$ ($m\rightarrow n$), and $p_{k}^{n}$
is the transmit signal power over the $n$-th cable pair at $k$-th
frequency band such that
\begin{equation}
p_{k}^{n}=\textrm{E}\big\{\left|\tilde{x}_{k}^{n}\right|^{2}\big\}.
\end{equation}
The power $p_{k}^{n}$ is optimized to minimize the crosstalk toward
the other lines ($n\neq m$), and the optimization problem for all
the cable pairs ($n$) and sub-carriers ($k$) can be stated as
\begin{align}
\ \underset{\mathbf{p}^{1},\ldots,\mathbf{p}^{N_{c}}}{\max} & \stackrel[n=1]{N_{c}}{\sum}R^{n}\nonumber \\
\textrm{s.t. } & P^{n}\leq P^{n,Tot},\,\,\,\,\,\,\,\,\,\,\,n=1...N_{c},\nonumber \\
 & 0\leq p_{k}^{n}\leq p_{k}^{n,\,mask},n=1...N_{c}\textrm{ and }k=1...N_{f},\label{eq:Power_Optimization_Problem-1}
\end{align}
where $R^{n}$ is the throughput of the $n$-th pair by summing the
contributions from $N_{f}$ sub-carriers with frequency spacing $f_{s}=22$
MHz (as in Fig. \ref{fig:Cat6_characteristics}):
\begin{equation}
R^{n}=f_{s}\underset{k}{\sum}\log_{2}\left(1+\text{SINR}_{k}^{n}\right).
\end{equation}
$P^{n}=\stackrel[k=1]{N_{f}}{\sum}p_{k}^{n}$ is the sum power of
all the sub carriers for line $n$, and it is constrained by the maximum
transmit power of the cable amplifier $P^{n,Tot}$. The maximum transmit
power per subcarrier is also constrained to \textbf{$p_{k}^{n,\,mask}$}.
The required transmit power on $k$-th subcarrier for all the cable
pairs is \cite{8,9}

\begin{equation}
\mathbf{p}_{k}=\left(\mathbf{D}_{k}-\boldsymbol{\Lambda}_{k}\mathbf{A}_{k}\right)^{-1}\boldsymbol{\Lambda}_{k}\boldsymbol{\sigma}_{k},
\end{equation}
where
\begin{align}
\mathbf{D}_{k} & =\textrm{diag}\left[\vert h_{k}^{1,1}\vert^{2},\,\vert h_{k}^{2,2}\vert^{2},\,\dots,\,\vert h_{k}^{N_{c},N_{c}}\vert^{2}\right]\\
\boldsymbol{\Lambda}_{k} & =\textrm{diag}\left[\text{SINR}_{k}^{1},\,\text{SINR}_{k}^{2\qquad},\,\dots,\,\text{SINR}_{k}^{N_{c}}\right]\label{eq:SNIR}\\
\left[\mathbf{A}_{k}\right]_{n,m} & =\left\{ \begin{array}{c}
0\qquad\;\;\;\;\;\quad m=n\\
\left|h_{k}^{n,m}\right|^{2}\;\;\;\quad m\neq n
\end{array}\right.
\end{align}
and the required SINR in (\ref{eq:SNIR}) are selected iteratively
for different M-QAM constellations to guarantee that power allocation
is positive valued. The numerical values here are derived from the
LTE specifications \cite{10}. The spectrum balancing algorithm in
\cite{8} provides an efficient solution for the optimization problem
in (\ref{eq:Power_Optimization_Problem-1}), in the form of an amplifier
gain matrix $\mathbf{B}\in\mathbb{R}^{N_{a}\times N_{a}}$. Such matrix
is block diagonal as in (\ref{eq:ref_channel_diag}): $\mathbf{B}=\text{diag}(\mathbf{B}_{1},\dots,\mathbf{B}_{N_{f}})$
and the $N_{c}\times N_{c}$ amplifier gain matrix $\mathbf{B}_{k}$
for $k$-th subcarrier is 
\begin{equation}
\mathbf{B}_{k}=\textrm{diag}\left[\sqrt{p_{k}^{1}},\,\sqrt{p_{k}^{2}},\,\dots,\,\sqrt{p_{k}^{N_{c}}}\right].
\end{equation}
To comply with the air-link specification for LTE, the spectral efficiency
is upper-limited to 8 bps/Hz, corresponding to the maximum constellation
of the evolution of LTE that supports up to 256-QAM.

\section{Numerical Results\label{sec:Numerical Results}}

\begin{figure*}[t]
\noindent \begin{centering}
\includegraphics[width=1\columnwidth]{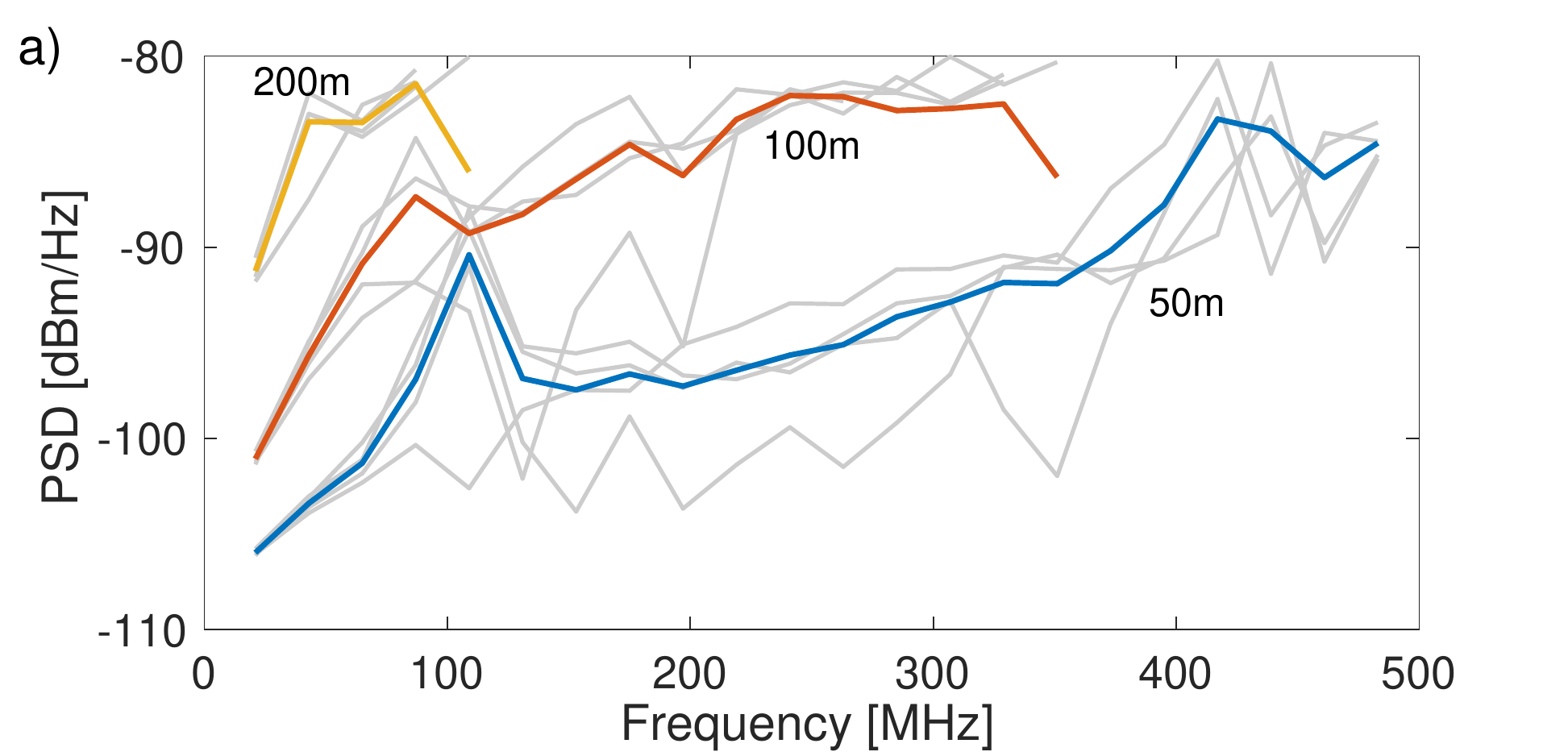}\includegraphics[width=1\columnwidth]{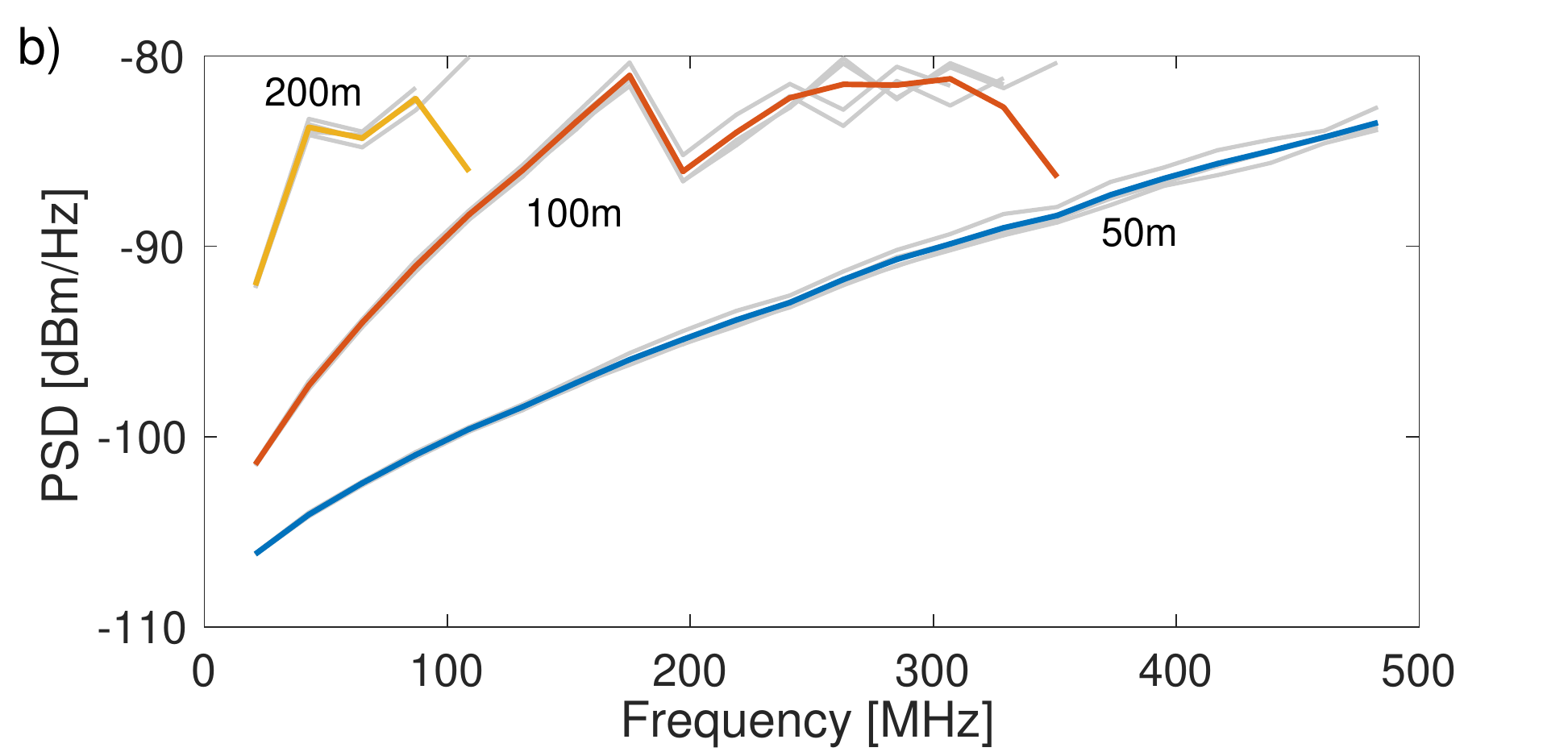}
\par\end{centering}
\begin{centering}
\includegraphics[width=1\columnwidth]{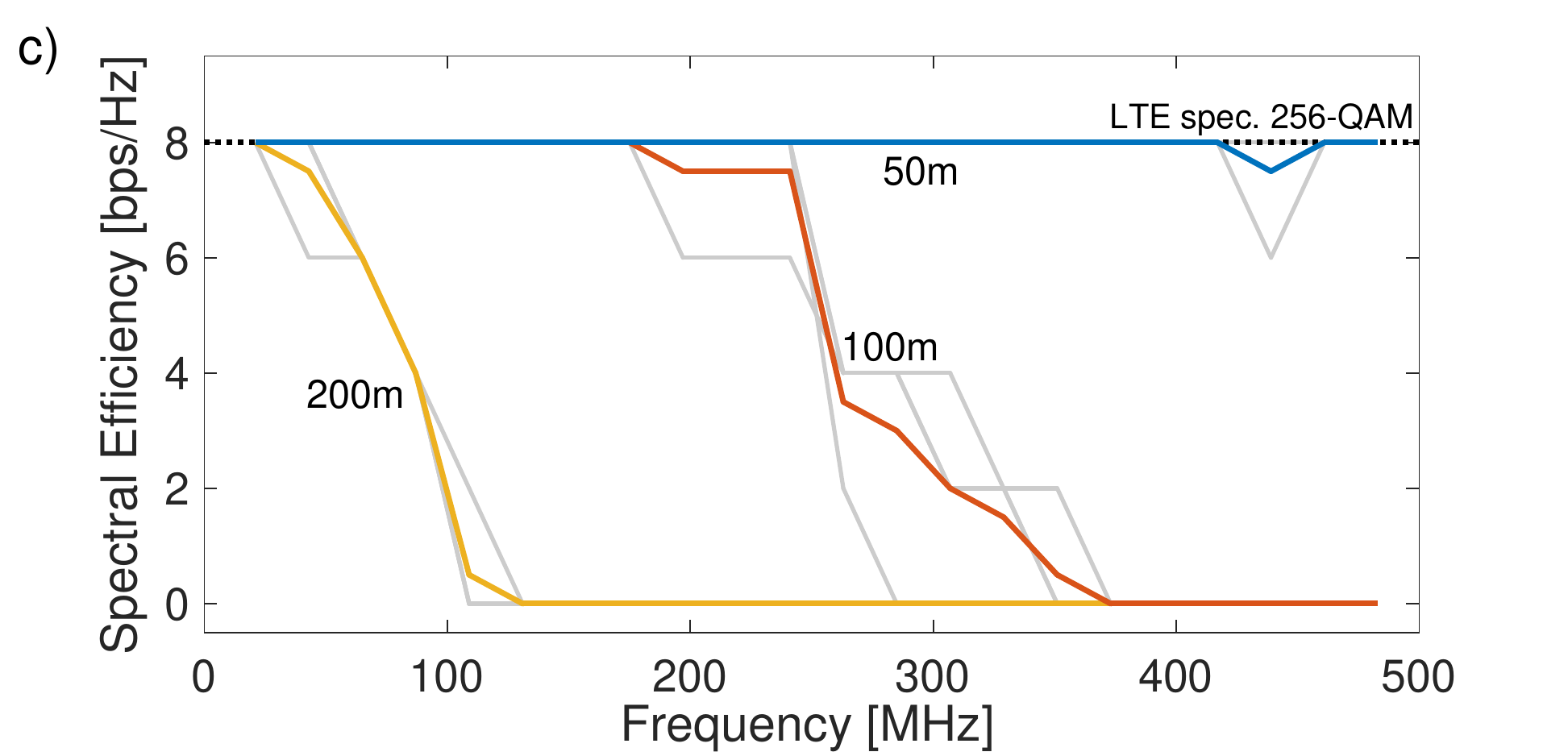}\includegraphics[width=1\columnwidth]{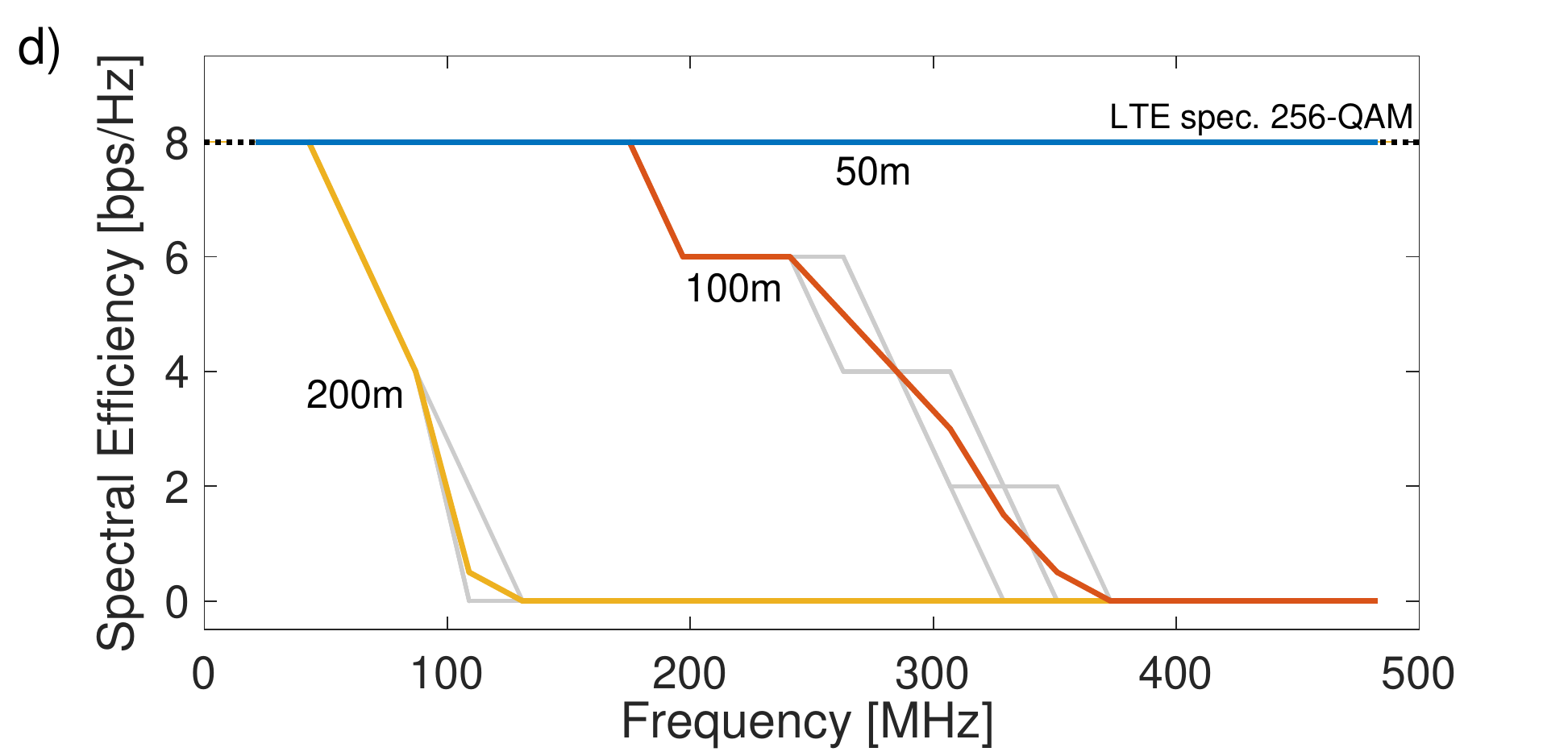}
\par\end{centering}
\noindent \centering{}\caption{Transmit PSD (top) and average spectral efficiency vs frequency (bottom),
with OSB and with (right) or without (left) cable crosstalk compensation
for different lengths, Cat-5 cable.\label{fig:Link-budget-with}}
\end{figure*}

Simulation results for different cable types and lengths, based on
the input parameters summarized in Table \ref{tab:SB-Parameters-for},
are presented here to show the effectiveness of the proposed fronthauling
architecture. In particular, the number of independent 20-MHz LTE
signals (assumed to be equal to the number of antennas $N_{a,max}$)
available on the cable channel is evaluated by considering the modulation
parameters from LTE specifications \cite{10}. 
\begin{table}[t]
\vspace{-5pt}

\includegraphics[width=0.78\columnwidth]{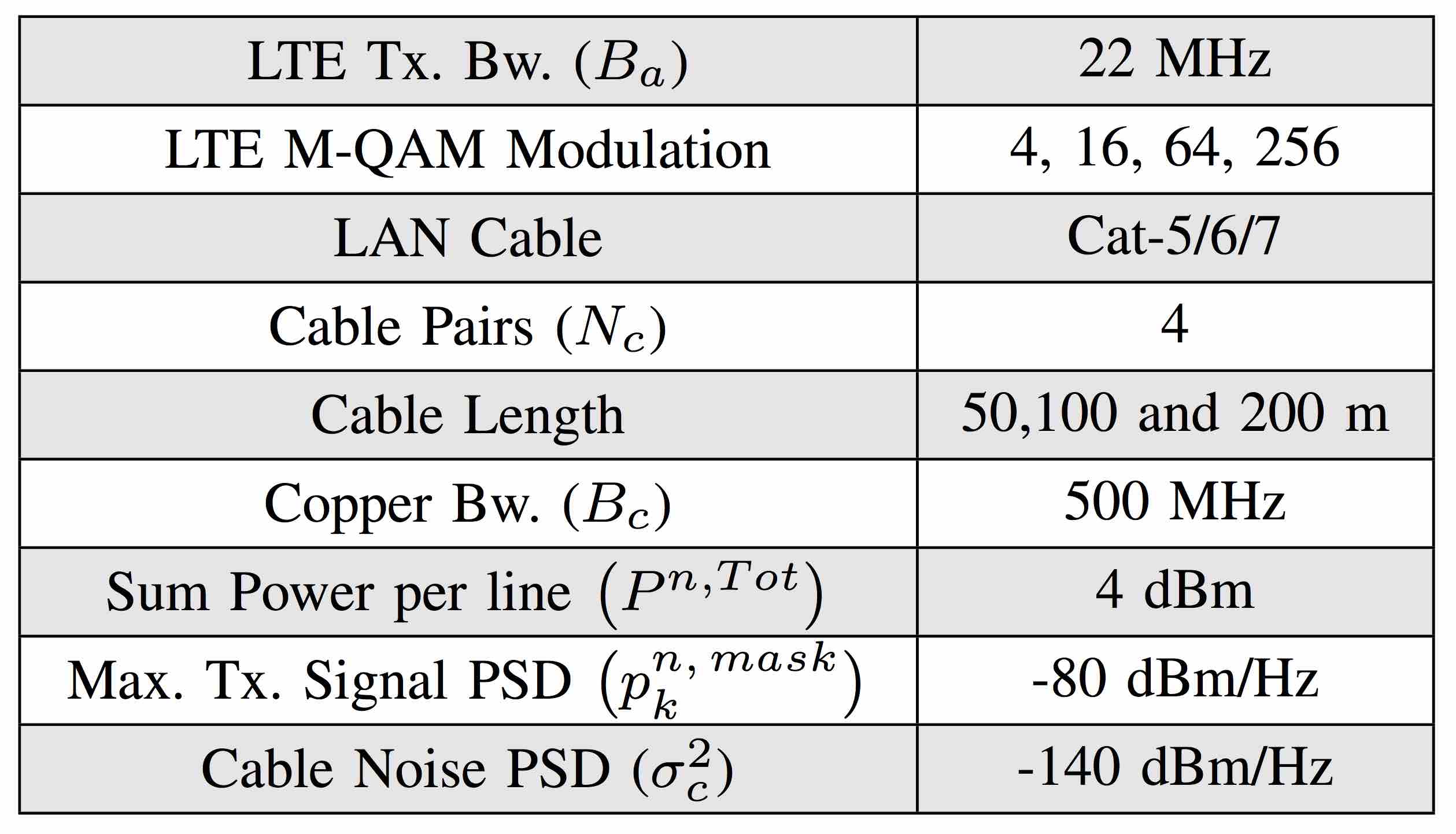}
\begin{centering}
\caption{System Parameters for MIMO-RoC. \label{tab:SB-Parameters-for} }
\par\end{centering}
\vspace{-15pt}
\end{table}
The copper bandwidth $B_{c}$ is limited here up to 500 MHz even if
the usage can be expanded up to 1 GHz and beyond, as foreseen for
future broadband access networks \cite{effenberger2016future}. 

The results of OSB as defined in (\ref{eq:Power_Optimization_Problem-1})
are shown in Fig. \ref{fig:Link-budget-with} for a Cat-5 cable, and
reported in Tables \ref{tab:Link-budget-summary-1-1}, \ref{tab:Link-budget-summary-1}
for Cat-5/6/7. In order to compare the OSB for RoC with digital pre-compensation
of FEXT at RAU, the performance of conventional Tomlinson-Harashima
precoding (THP) for FEXT compensation in multi-pair copper cables
\cite{hekrdla2015ordered} is shown as reference in Fig. \ref{fig:Link-budget-with}b,
d, and in Table \ref{tab:Link-budget-summary-1}. In particular, the
averaged (over the $N_{c}$ pairs) OSB transmit powers over the 22-MHz
bins are in the upper part of Fig. \ref{fig:Link-budget-with}, while
the powers for each of the 4 twisted pairs are in gray lines. The
corresponding spectral efficiency that corresponds to the M-QAM modulations
of LTE according to the specifications is in the lower part of the
same figure. 
\begin{table}[h]
\vspace{-10pt}

\centering\includegraphics[width=0.95\columnwidth]{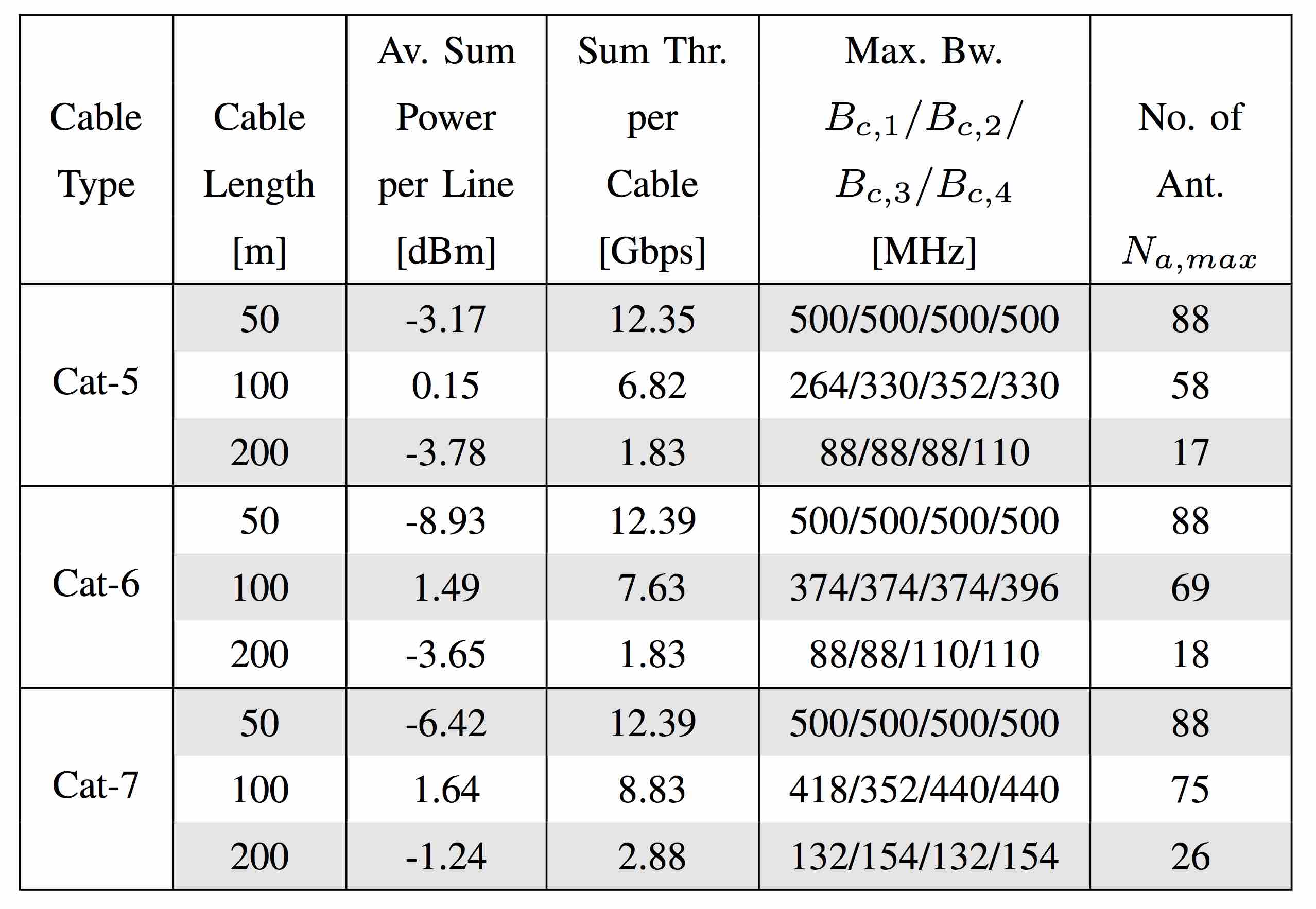}

\vspace{-4pt}

\begin{centering}
\caption{Cable resources and OSB parameters without cable crosstalk compensation.
\label{tab:Link-budget-summary-1-1} }
\par\end{centering}
\vspace{-15pt}
\end{table}
The power required for any given modulation scheme increases with
cable impairments, and consequently with frequency and length.

\begin{table}[h]
\vspace{3pt}

\centering\includegraphics[width=0.95\columnwidth]{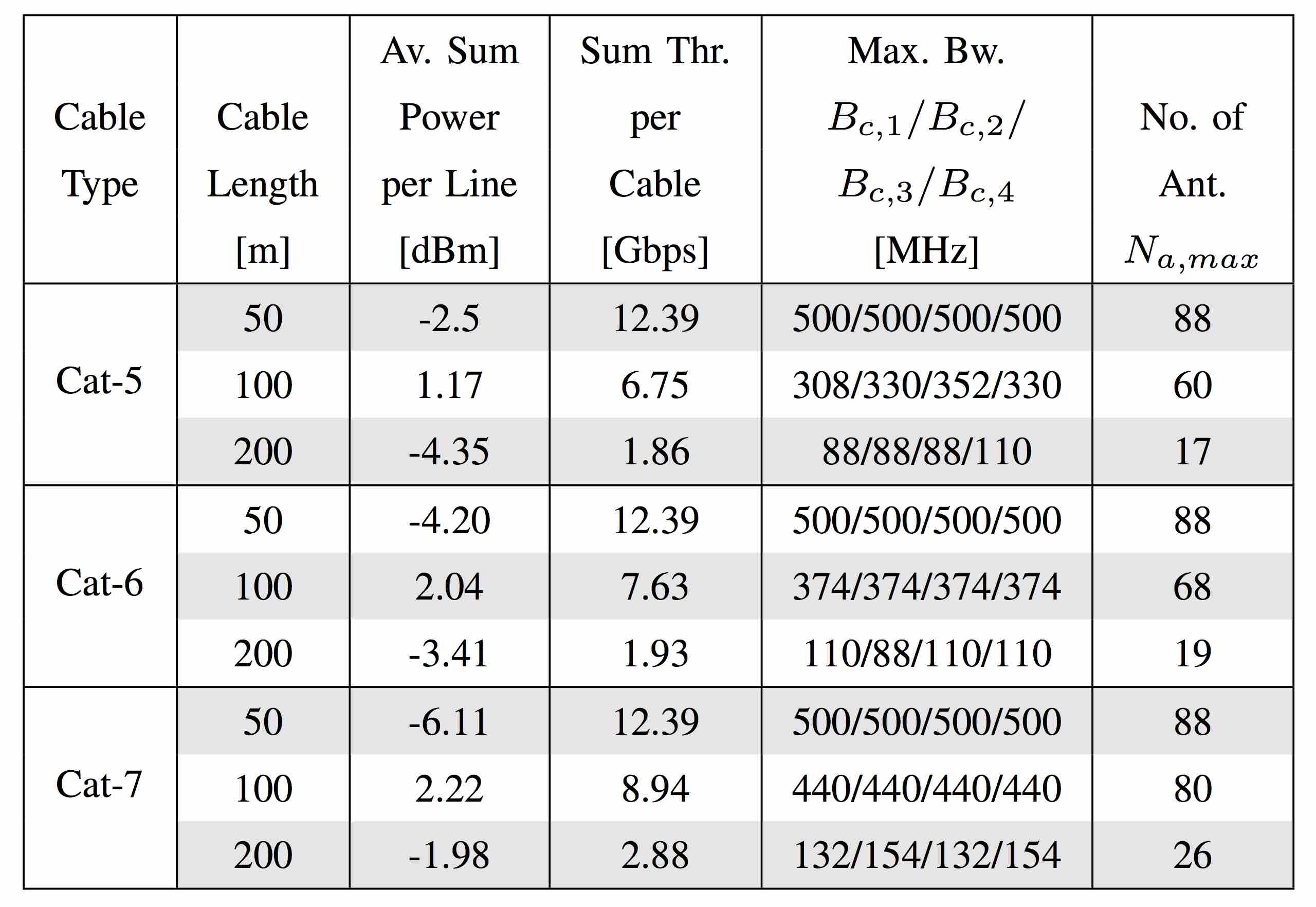}

\vspace{-4pt}

\begin{centering}
\caption{Cable resources and OSB parameters with cable crosstalk compensation.
\label{tab:Link-budget-summary-1} }
\par\end{centering}
\vspace{-15pt}
\end{table}

\begin{figure}[h]
\noindent \begin{centering}
\includegraphics[width=1\columnwidth]{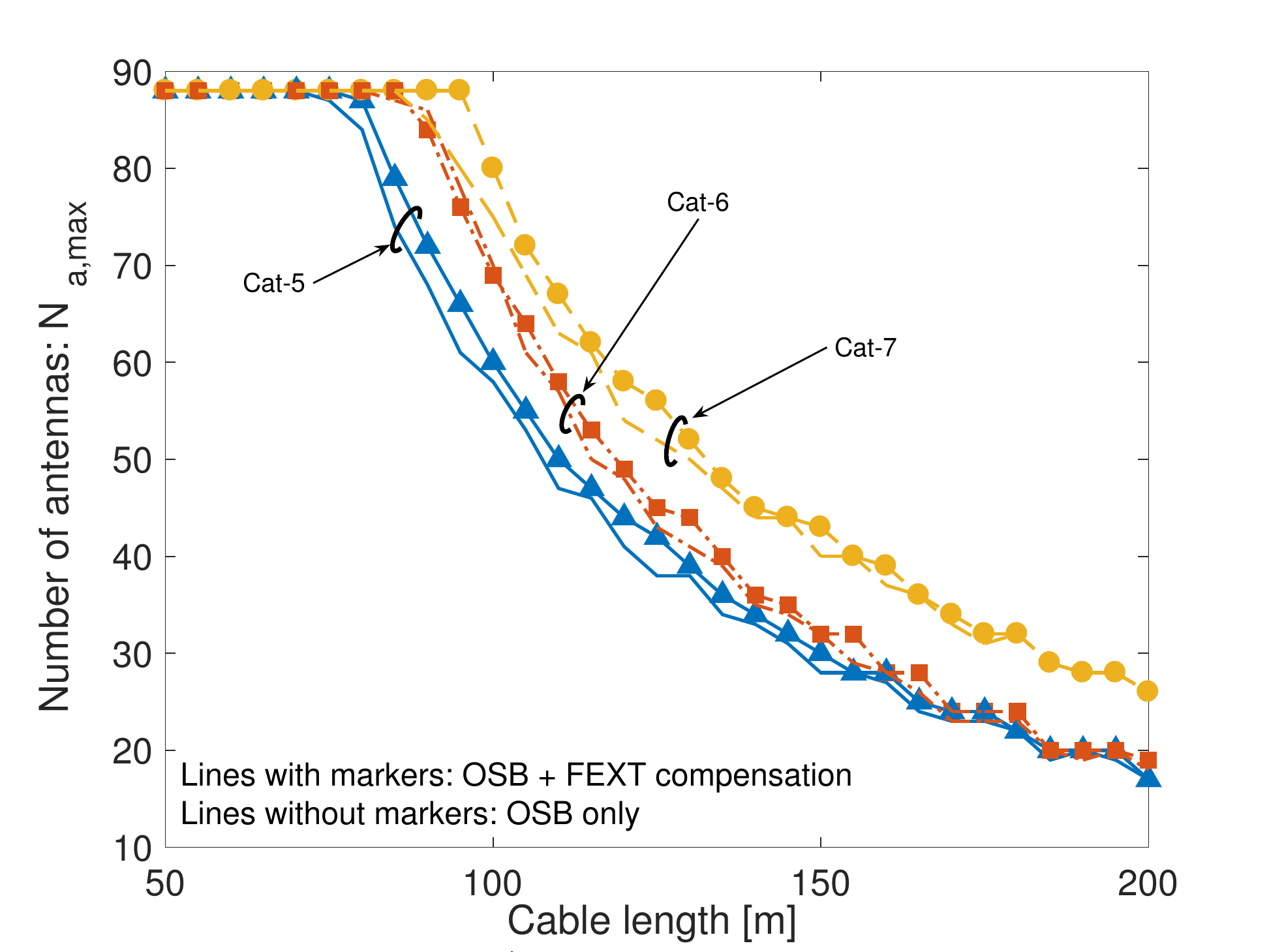}
\par\end{centering}
\vspace{-12pt}

\noindent \begin{centering}
\caption{Number of antennas vs cables length for Cat-5/6/7 with OSB and with
or without cable crosstalk cancellation\label{fig:Link-budget-with-1}}
\par\end{centering}
\vspace{-12pt}
\end{figure}
We observe that the 256-QAM (spectral efficiency of 8 bps/Hz) is guaranteed
for the entire 500-MHz bandwidth when using the 50-m cables. Increasing
the cable length the 256-QAM requirements can be fulfilled over a
reduced bandwidth up to approx. 175 MHz and 25 MHz for 100 and 200-m
cables. Comparison with digital FEXT compensation shows a modest improvement
compared to the cost. The maximum useful bandwidth for transmission
on each pair ($B_{c,\ell}$, second to last column in Tables \ref{tab:Link-budget-summary-1-1},
\ref{tab:Link-budget-summary-1}) is obtained from Fig. \ref{fig:Link-budget-with}c,d
as the spectrum portion corresponding to non-zero spectral efficiency
and the total number of antennas is computed for the useful transmission
bandwidth as in (\ref{eq:Cable_Capacity-1}). It is to be noticed
that the average sum power per line is always lower than 4dBm, that
is the limit of commercially available power driver for twisted-pairs.
As expected, the maximum throughput is achieved by Cat-7 cables, due
to its higher FEXT protection capabilities.

The total number of allocable independent 20-MHz LTE channels that
corresponds to the number of antennas $N_{a,max}$ versus cable length
is in Fig. \ref{fig:Link-budget-with-1} for all cable types and considering
also the impact of FEXT processing at the RAU as design parameter.
Up to around 75 m, any cable (regardless the use of FEXT compensation)
allows to serve the maximum number of antennas compatible with 500-MHz
bandwidth that is $88\simeq4\times500\text{ MHz}/22\text{ MHz}$ (i.e.,
22 LTE channels on each pair), which is limited by the 500-MHz bandwidth
for reliable cable measurements. Distance of 100 m on a LAN cable
is considered as a reference, and at least 55 independent LTE channels
(or independent antennas) can be transported using MIMO-RoC on Cat-5,
thus proving the effectiveness of MIMO-RoC fronthauling for the last
100 m of the C-RAN architecture. It is important to notice that the
beneficial effects of crosstalk compensation are modest, and can be
appreciated only for lower cable lengths where the interference dominates
over noise. This is a strong argument in favor of the usage of all-analog
processing of judicious power allocation in all-analog MIMO-RoC fronthauling
to minimize any latency of the I/Q signal transport.

\section{Concluding Remarks \label{sec:Conclusion}}

In this paper we consider the implementation of fully-analog fronthauling
based on the RoC paradigm that is based on the use of LAN cables,
already largely deployed in buildings and enterprises. In particular,
we prove the great potential provided by multi-pair LAN cables, evaluated
in terms of number of antennas that can be served for different cable
lengths and types. By numerical evaluation, we prove that in a 50-m
Cat-5 cable the maximum number of antennas (or 20-MHz LTE channels)
is 88, and this value is limited by the cable bandwidth of 500 MHz
considered here. In particular, at least 60 antennas can be served
by a cable of practical length (100 m), thus enabling the design of
MIMO-RoC fronthauling to serve a RAU with a large number of antenna
elements. Pushing cable bandwidth up to 1 GHz and beyond, it can be
shown that the number of allocable RAT channels rises up to approx
120. The proposed radio-MIMO over cable-MIMO allows the joint exploitation
of space and frequency multiplexing on both cable (multiple pairs)
and air (multiple antennas). Hence a further degree of freedom in
the design of the proposed architecture is given by the mapping between
the space-frequency channels defined by antennas and radio spectrum
onto the cable-pairs and cable spectrum. This promising activity is
planned as future work, together with a thorough cost-benefit comparison
with RoF.

\bibliographystyle{ieeetr}

\end{document}